# Cross-sectional profile of photocarrier mobility in thin film solar cells via multimolecular recombination and charge extraction by linearly increasing voltage (cs-p-CELIV)

Noah B. Stocek, Miguel J. Young, Reg Bauld, Tianhao Ouyang, and Giovanni Fanchini*

*Department of Physics & Astronomy, The University of Western Ontario, 1151 Richmond St. London ON, N6A 3K7, Canada*

**Abstract -** The ability to spatially resolve the carrier mobility profile along the cross section of micrometer-thin solar cells is vital, both for fundamental studies in photovoltaics and as a quality control for reproducibly obtaining high conversion efficiencies in commercial solar cell modules. Presently, no technique capable of such an endeavor is available to the best of our knowledge. Here, we introduce a novel method capable of profiling the carrier mobility along the *z*-axis in thin-film photovoltaics. Our setup is based on the integration of photogenerated charge extraction by linearly increasing voltage (p-CELIV) with a scanning confocal optical microscope (SCOM) towards a confocal and cross-sectional p-CELIV (cs-p-CELIV) system. As monomolecular recombination of excess carriers is the most frequent radiative pathway for electrons and holes in solar cells at low power density of illumination, while multimolecular recombination dominates at high power, enhanced multimolecular recombination occurs at the SCOM focal plane. Thus, the cs-p-CELIV signal provides enhanced information on the mobility of all of the cross-sectional layers, except the focal plane. By scanning the focal plane along the *z*-axis, the mobility profile can be derived. To demonstrate our technique, we use it to investigate the carrier mobility in three hydrogenated amorphous silicon (a-Si:H) solar cells. The mobility profiles obtained by cs-p-CELIV correlate well with well-known depletion layers effects, as well as the H content profile in a-Si:H, which is measured independently. Our findings are in excellent agreement with models suggesting a critical role of Si-H bonding in locally determining the carrier mobility in a-Si:H.

---

* Corresponding author. E-mail: gfanchin@uwo.ca

# 1. Introduction

The design of advanced solar cells has led to tremendous efforts towards the development and characterization of high-quality semiconducting materials capable of converting sunlight into electrical power at low cost and high photoconversion efficiency.[1-3] A physical property of unique importance in solar energy materials and solar cells is the carrier drift mobility.[4,5] The mobility is a measure of the drift velocity at which charge carriers can diffuse in solids when acted on by an electric field.[6] The term carrier mobility refers in general to both electron and hole mobilities, which in semiconducting materials may be very different, and may also differ in the dark and upon illumination. The photoconversion efficiency of solar cells can be dramatically different depending on whether there are many carriers with low mobility or few carriers with high mobility in their active layer.[7] Almost always, higher mobility leads to better device performance,[8] all other device properties being equal. Development of laboratory techniques capable of measuring and imaging the carrier mobility in photo-active semiconducting layers is critical to improve the performance of thin-film solar cells incorporating these materials.

A well-established method for charge carrier mobility measurements relies on the Hall effect [9] in which a transversal voltage is created between two parallel faces of a sample at a constant magnetic field. A drawback of Hall effect mobility measurements is that these experiments require a very intense magnetic field in low-mobility photovoltaic devices, and they only carry information on the mobility of dark carriers. Differently from Hall effect experiments, mobility measurements in the time domain [10,11] are sensitive to the mobility of photoexcited carriers and do not require any magnetic field application. Time-domain photoexcitation techniques include time of flight (ToF) [10] and photo-conductivity recovery dynamics (PRD).[11] These techniques also have limitations, however. For example, PRD does not allow for mobility measurements independent of the electrical transport properties, but rather yields the mobility lifetime product, while ToF techniques offer reliable results only on relatively thick photoactive films, of a few μm or more.[12]

Photogenerated charge carrier extraction by linearly increasing voltage (p-CELIV) [13] is a time-domain mobility measurement technique that is applicable to devices with less restrictive thickness requirements than ToF.[12] Like ToF, p-CELIV is carried out by illuminating the sample with a nanosecond laser pulse to photoexcite a packet of charge carriers. Differently from ToF experiments, the voltage waveform used to induce a p-CELIV current transient is linearly increasing over time. A linear voltage ramp generates a significantly lower photocarrier drift velocity than a constant square waveform, as in ToF. Thus, at any device thickness and carrier mobility, the transient current maximum occurs at longer extraction times than ToF. This indicates that p-CELIV is capable of mobility measurements through thinner and less optically dense devices, typically down to 100 nm in hydrogenated amorphous silicon (a-Si:H) solar cells.[14]

A significant limitation of state-of-the-art time-domain photoexcitation techniques is that they are normally used only as one-data-point measurements, with no ability of profiling the mobility along the cross section of the active layer. While the suitability of p-CELIV to map the carrier mobility at micron-scale resolution along the *x-y* plane of organic photovoltaics has been reported,[15] demonstration of spatially resolved mobility scans along the *z*-axis requires optical resolution well below 1 μm, with the integration of p-CELIV with a scanning microscope, or *ad hoc* physical understanding of the carrier drift and recombination processes.

Here, we introduce a novel cross-sectional technique based on p-CELIV, which is capable of scanning the carrier mobility profile of a solar cell along the *z*-axis. To this end, a p-CELIV apparatus will be integrated with a scanning confocal optical microscope (SCOM) to induce strong light power density at the SCOM focal plane, but significantly lower illumination intensities above and below. Multimolecular recombination can be defined as the recombination of a carrier (electron or hole) with an opposite carrier (hole or electron, respectively) that is not the same one with which it has been photogenerated. Thus, multimolecular recombination occurs only when high enough concentration of electron-hole pairs is locally generated due to high enough photon flux. Because of this, enhanced recombination of excess carriers in our system will occur at the focal plane due to multimolecular recombination processes [16,17] leading to diminished charge extraction from such a plane, and additional information on the device can be simultaneously acquired via confocal optical microscopy. This will allow for the returning of an array of p-CELIV signals by scanning one focal plane at a time, with each signal containing enhanced information on the mobility of all of the cross-sectional layers, with the exception of the focal plane, thus effectively implementing a cross-sectional p-CELIV (cs-p-CELIV) technique. Cross-sectional mobility profiles along the *z*-axis will thus be compiled through a specially developed algorithm that includes the discretization of the active layer in multiple slabs and considers multimolecular radiative recombination near the focal plane.

## 2. Experimental

### 2.1. Cs-p-CELIV apparatus

The p-CELIV and cs-p-CELIV measurements presented in this study were recorded using a Meshtel Intelite Q-switched laser at $\lambda = 527$ nm wavelength, $t_p = 8$ ns pulse duration, 0.01-30 kHz repetition rate, and 60 µJ pulse energy. As the used 527 nm wavelength falls within the fundamental optical absorption region of a-Si:H, where no interference fringes are detectable, for example by UV-visible spectroscopy, thin film interference can be neglected. The laser was connected to a WITec Alpha300 SCOM [Figure 1(a)] by a step-index multimode optical fiber (Thorlabs Inc., 0.22 numerical aperture and 200 μm core diameter) which was wired to the SCOM pinhole. The resulting beam had a 2.8 $10^{-5}$ cm$^2$ spot size at the focal plane, which allows for a high enough energy density (2.1 J cm$^{-2}$ per pulse) to ensure strong multimolecular recombination in a-Si:H [16]. As the points spread function (PSF) decreases rapidly outside of the focal plane in a confocal microscope, this also ensures that the laser energy density just outside the focal plane is on the order of mJ/cm$^2$. The optical setup used in this experiment allowed for 100 nm thick slices of multimolecular regions. The SCOM is equipped with a computer-controlled piezo-scanner with 20-mm *x-y* travel range, 30-mm *z*-axis travel range, 0.1 nm *z*-axis resolution and ±2nm repeatability, on which the sample was positioned, as shown in Figure 1(b). The SCOM is attached to a Hamamatsu H8259 photomultiplier operating in photon-counting mode.

To record cs-p-CELIV data at each (*x, z*) point of the device active layer, the end-of-pixel signal from the WITec Alpha 300 SCOM control unit is used to trigger the Q-switched laser. At a time lag $t_D$ = 200 ns, obtained through a tunable delay circuit, the triggering signal reaches a Stanford Research DS345 arbitrary function generator, which is set to produce a sequence of linearly increasing voltage ramps at B = 2.45 $10^5$ V s$^{-1}$ slope, $t_{ramp}$ = 40 μs duration, $U_0$ = -1 V offset, and 100 μs repetition rate. A delay time of 200 ns was used as it allows for photo-generated charge carriers to thermalize to equilibrium in bulk measurements. The function generator is connected to a bridge circuit on which a 1000-Ω resistor is placed in parallel to the solar cell under investigation, as depicted in Figure 1(a). This value of bridge resistance led to a suitable RC-time of the measurement circuit for both bulk and cross-sectional measurements, so capacitive effects within the signal can be disregarded. A Hantek 5200A digital scanning oscilloscope operated through a custom-made LabView$^{TM}$ routine is connected to the photovoltaic device and is used to acquire the data. An ASCII file containing the cs-p-CELIV signal is recorded for each (*x, z*) pixel and stored in a personal computer for further analysis, as will be discussed in Sect. 3. For comparison, a macroscopic p-CELIV measurement (i.e. without using the SCOM) was carried out as reported in the literature [18-20] at $t_D$ = 200 ns, B = 2.45 $10^5$ V s$^{-1}$ slope, $t_{ramp}$ = 40 µs, $U_0$ = -1 V, 100 µs repetition rate, and P = 1.9 $10^{-3}$ J cm$^{-2}$ per pulse energy density at the sample surface.

## *2.2. Solar cells used to validate this study*

Three solar cells were used to validate our cross-sectional photo-CELIV measurement apparatus. The first, labelled sample 1, is a Sanyo AM-1411 hydrogenated amorphous silicon (a-Si:H) commercial thin-film module, composed of three 29.6 x 11.8 mm devices connected in series. The second, labelled sample 2, is an Uberhaus a-Si:H commercial module, composed of three 20 x 10 mm devices, again connected in series. Sample 3 is also an Uberhaus a-Si:H solar cell, of higher efficiency than sample 2, and is composed of three 30 x 10 mm devices in series. It is anticipated that the active layer of sample 3 is thinner than that of samples 1 and 2, due to its higher efficiency. For this study, a single device was detached from each module by cutting the glass substrate and has been electrically connected to the p-CELIV apparatus.

The two remaining devices from the module of sample 1 have been used for *post-mortem* characterization, including scanning electron microscopy (SEM), elastic recoil detection analysis (ERDA), and secondary ion mass spectroscopy (SIMS). This solar cell has been chosen because its specifications are representative of a large class of a-Si:H photovoltaics. As depicted in Figure 2(a) these devices have a *p-i-n* architecture [21] with an intrinsic a-Si:H region of about 1-2 µm thickness sandwiched between *p*-doped and *n*-doped regions that are significantly thinner, 30 nm or less.[21] Samples 2 and 3 are of the same architecture but with intrinsic layers of different thicknesses. 2. As the thickness of all these solar cells (up to a few µm) is much smaller than the side length of the electrodes (several mm) and the beam was centered on the surface of the top electrode, the CELIV assumption of linear electric field remains a very reasonable assumption.

*Post-mortem* cross-sectional analysis of a device from the same module was carried out on a Zeiss 1540 XB SEM, equipped with an energy dispersive X-ray (EDX) detector and a focused ion beam (FIB) apparatus. A trench was cut by Ga-ion FIB milling down to 4 μm, an in-depth penetration that was expected to be higher than the device thickness. The sample was subsequently tilted on the SEM stage, and a cross-sectional image [Figure 2(b)] was thus recorded. In this way, we were able to confirm that the active layer thickness is 1.2±0.1 μm, with an 800±100 nm thick aluminum top electrode, and an indium tin oxide (ITO) cathode of 1.0±0.1 μm grown on a glass substrate. These figures are consistent with the typical design of a-Si:H *p-i-n* photovoltaics.[21] The assignment of each layer to Al, a-Si:H, ITO and glass was confirmed by EDX imaging, recorded for Al, Si, and Sn along the same cross section of the SEM image. Al and Si content maps are reported in Figures 2(c) and (d), respectively, and help to define the electrodes, active layer, and glass substrate, to unambiguously identify them by SCOM during cs-p-CELIV measurements.

The hydrogen profile in the a-Si:H solar-cell active layer was investigated using elastic recoil detection analysis (ERDA) and dynamic secondary ion mass spectrometry (SIMS) to establish a correlation between charge-carrier mobility and H content at each point along the cross section. ERDA was performed at Interface Science Western on a High Voltage Engineering Europa Tandetron ion accelerator, using a 1.7 MeV $^4$He$^+$ ion beam at 75° incidence angle with respect to the sample surface. Recoiled hydrogen was detected at a collection angle of 35° on a silicon diode retrofitted with a 700-channel analyzer and protected with a range foil to eliminate forward scattered ions from the incoming beamline. ERDA measurements were modelled using the SIMNRA simulation package.[22] In these simulations, the total thickness of the active layer was assumed to be 1.2 μm in accordance with SEM measurements, and the active layer was divided into 16 slabs at a constant thickness of 75 nm, for which different H contents were assumed. SIMS data were recorded on a Cameca IMS-6F spectrometer from a 180 μm-by-180 μm raster-scanned sample area. Oxygen ions were accelerated at 10 keV, which resulted in the erosion of the a-Si:H layer at a controlled rate, with the emission of secondary ions. Such ions were subject to a 4.5-keV accelerating voltage and analyzed in a magnetic sector mass spectrometer. This allows for the quantification of the secondary ion beam elements, in particular hydrogen, at different depths within the solar cell active layer, until the active layer was fully eroded and In from the ITO cathode was detected. A calibrated reference sample, with hydrogen implanted into a Si wafer, was used for H content calibration during the SIMS measurements.

As a final method of characterization, $j-V$ curves were measured for all 3 of the samples under a Newport AM 1.5 solar simulator, following the procedure outlined by the national renewable energy laboratory (NREL).[23] From this, the short-circuit current density, open-circuit voltage, maximum power, and efficiency were obtained, and are shown in Table 1. A Keithley 2400 power supply was interfaced using a Matlab$^{TM}$ routine to scan over a specified voltage range and record current, and an NREL certified reference solar cell was used to calibrate an Oriel Instruments 6253 150-W xenon arc lamp, light from which passed through an AM 1.5G filter and was incident on the sample's surface. The intensity and position of the xenon lamp were adjusted until all reference cell parameters were within 5% of the NREL certified values. The ratio of our measured values to the certified values was used as a calibration factor (which was between 0.95 and 1.05 for all parameters) for subsequent measurements on the three samples. These results show that both samples 1 and 2 are of lower efficiency, while sample 3 has a higher efficiency, close to 10%. This allows for testing of our technique on devices with a variety of properties.

## 3. Discretized models for p-CELIV data analysis

Models used for single-data point p-CELIV analysis [14,24] typically rely on a continuous medium approximation of the active layer. They are, therefore, difficult to be generalized to analyze cs-p-CELIV experiments. In these models, [14,24] the active layer is assumed to be uniform along the device cross section. Under the additional assumptions of very different mobility for majority and minority carriers and low conductivity of the active layer, the majority carrier mobility can be calculated [14,24] from the signal transient as

$$\bar{\mu} = \frac{2D^2}{3Bt_m^2(1+0.36j_m/j_0)}, \quad (1)$$

where $D$ is the active layer thickness, $j_m$ is the maximum transient photocurrent that occurs at a time $t_m$ after the transient started, $j_0$ is the constant current through the capacitive device due to a voltage ramp $V(t) = Bt$, with the voltage ramp starting at a time $t = 0$ and the laser pulsing at $t = t_D$ (Figure 1a). This model, shown in Figure 3a, assumes that, at any $t$, a region of the active layer at $0 < z < l(t)$ only contains minority photocarriers (majority photocarriers have already been extracted) while both photocarrier types are present at $l(t) < z < D$. Thus, the extraction layer boundary $l(t)$ drifts across the device at a speed corresponding to the majority photocarrier drift velocity, $v_d(t) = \bar{\mu} E(t)$, where the electric field $E(t)$ linearly depends on $t$ the same way as $V(t)$, and is assumed to be independent of $z$. Thus, both $v_d(t)$ and $\bar{\mu}$ represent averages over the device $z$-axis. Because equation (1) offers accurate majority carrier mobility estimates only in devices that are relatively uniform along the $z$-axis cross section, a model for interpreting p-CELIV and cs-p-CELIV data in nonuniform devices is required in the present study.

To produce a model suitable to interpret cs-p-CELIV experiments in which the mobility of majority photocarriers varies along the $z$-axis, we will here develop discrete models of photocarrier drifts in which the device's active layer is divided into $k = 1, ..., M$ slabs of identical thickness $D/M$ ($M \gg 1$). Each homogeneous slab may have a different density of photogenerated carriers ($n_k$) and carrier mobility ($\mu_k$). We will first prove (sect 3.1) that, under the assumption of non-focal illumination and $\mu_k \approx \bar{\mu}$ for any $k$, our discrete-layer model yields the same results as the continuous medium approximation and, therefore, equation (1). We will then explore the effects of focal illumination at different levels $k$ of the active layer. For the sake of simplicity, we will initially confine ourselves (sect 3.2) to highly focal illumination of semitransparent devices.

The semitransparent device approximation will be shown to be equivalent to assume that the same amount of laser power ($W_0$) is absorbed at any layer $k$ of the device. Highly focal illumination is tantamount to assume, in a "Dirac delta" approximation, that light is highly focussed only at one layer within the device (i.e., at $k = f$) where the illuminated area $A_k$ is much smaller than anywhere else along the cross section, and $A_f \ll A_{k \neq f} = A_0$. Therefore, while the *absolute* absorbed laser power (i.e.: $W_0$, in watt) is independent of $k$, the corresponding power *density* (i.e.: $P_k$, in watt *per unit volume*) is much larger at the focal plane: $P_f \gg P_{k \neq f} = P_0$. This means that, in a time unit, the same *absolute* number of photons (electron-hole pairs) are absorbed (photogenerated) at $k = f$ and any $k \neq f$. Electron-hole pairs are, however, produced much closer from each other at $k = f$, where the illuminated area is much smaller. This leads to a higher probability of multimolecular recombination at the focal plane, which is superlinear with $P_k$ [25, 26] and is, therefore, more likely to occur whereas the distance between neighboring electron-hole pairs decreases. Due to the additional effects of multimolecular recombination at the active layer, a smaller *absolute* number of charge carriers ($N_f$) is dissociated per unit time at $k = f$ than any other more broadly illuminated layer $k \neq f$ along the cross section. Thus, $N_f < N_{k \neq f} = N_0$ even if the carrier *density* is higher in the much smaller area $A_f$ at the focal plane, and $n_f \gg n_{k \neq f} = n_0$. This is extremely significant because it enables a layer-selective detection of the p-CELIV signal that, combined with layer-selective illumination from the SCOM, implements an unparalleled form of confocal microscopy for cross-sectional charge-carrier mobility measurements.

Once the essential ingredients of our model are established, we will extend it in sect 3.3 beyond the semitransparent device and highly focal illumination approximations. Specifically, we will model the laser power release along the device *z*-axis accordingly to Beer-Lambert law. As far as the "Dirac delta" approximation is concerned, we will relax it by assuming that *f* represents a set of slabs where the focal region will be broadened from a Dirac delta to a Gaussian-shaped distribution. With these generalizations, our model will be shown to suit well for cross-sectional photocarrier mobility estimates in a host of a-Si:H devices with very different properties.

### *3.1. Non-focal illumination of the active layer*

Figure 3a depicts a discretized p-CELIV model, where the active layer is divided into *M* slabs of identical thickness, the total photocurrent extracted at $t = t_k$ during the transient is the sum over all of the infinitesimal contributions from the carrier-containing slabs at $i > k$. We define $\tau_k$ as the time required for photogenerated carriers to transit through the *k*-th slab. Under a low conductivity assumption, the electric field redistribution is neglected, and only the applied ramp is considered for the field within each slab. Thus, the sum of currents extracted over all of the slabs can be expressed either by an integral or a discrete summation:

$$j_k = \int_{l(t)}^{D} N(z) e \mu(z) E(t_k) \frac{dz}{D} = e(N_{k+1}\mu_{k+1} + \ldots + N_M \mu_M) \frac{B t_k}{D}, \qquad (2)$$

where $N(z)$ and $N_k$ are the total number of majority photocarriers extracted at each level *z* in the continuous and discrete medium, respectively, and *e* is the elementary charge.

The approach normally used for extracting the carrier mobility $\mu$ is to determine the maximum conditions, $t_m$ and $j_m$. [24] With this approach, the maximum of the photocurrent transient, derived by differentiating $j_k$ over time in equation (2), is normalized by the constant dark current through the capacitive device leading, for example, to equation (1). By differentiating equation (2), the slope of the current transient at a generic time $t_k$ takes the form:

$$\Delta_k \approx \frac{j_{k+1}-j_k}{\tau_k} = \frac{eB}{D}[(N_{k+1}\mu_{k+1}+\ldots+N_M\mu_M)t_{k+1} - (N_k\mu_k+\ldots+N_M\mu_M)t_k]\frac{1}{\tau_k}. \quad (3)$$

Because the extracted photocurrent is zero at the onset of the linearly increasing voltage, to set $k = 1$ in equation (3), which leads to $t_k = \tau_k = \tau_1$ and $t_{k+1} \approx 2\tau_1$, simplifies such an equation into

$$\Delta_1 = \frac{eB}{D}(2N_1\mu_1 + N_2\mu_2 + \ldots + N_M\mu_M) \approx \frac{eB}{D}(N_1\mu_1 + \ldots + N_M\mu_M), \quad (4)$$

where the approximation holds for large $M$. Equation (4) can be used as an alternative to equation (1) to extract the carrier mobility from non-focal, single-data point, p-CELIV experiments. In a homogeneous discretized medium, $N_k \approx N_0$ and $\mu_k \approx \bar{\mu}$ are assumed to be constant and independent of $k$. The analytical expression for the p-CELIV current $j_k = j(t_k)$ [14] indicates a nearly linear increase of extracted photocurrent from zero to the maximum value. Therefore, the slope $\Delta_1 \approx j_m/t_m$ extracted from the experimental values can be used to determine $\bar{\mu}$.

While our data interpretation makes use of $k = 1$, a more general expression for $j_k$ can be worked out numerically under the assumption of constant $\bar{\mu}$ for comparison with p-CELIV models that take advantage of the continuous medium approximation. The instantaneous charge carrier drift velocity is determined by:

$$v_{d,k} = \frac{D}{\tau_k} = \mu_k E(t_k) = \mu_k \frac{Bt_k}{D}, \quad (5)$$

which can be used to find $\tau_k$ for the $k$-th layer by noticing that $t_k = \tau_1 + \ldots + \tau_k$. This is given by:

$$(\tau_1 + \ldots + \tau_k)\tau_k = \frac{D}{MB\mu_k}. \quad (6)$$

Starting from $\tau_1 \approx \sqrt{D/MB\mu_1}$, it is then possible to recursively calculate $\tau_k$ up to $k = M$ if a profile for $\mu_k$ is known, or $\mu_k \approx \bar{\mu}$ is posed.

Equation (6) is valid for any $\tau_k$. It depends on the field strength through voltage ramp slope, the active layer thickness, as well as the photocarrier mobility, with higher values of $t_k$ and/or $\mu_k$ leading to lower $\tau_k$, and more rapid charge extraction from the $k$-th layer. Evaluating equation (2) with the values of $\tau_k$ obtained from (6), and assuming a constant mobility $\bar{\mu}$ leads to results in line with the analytical expression (1). In agreement with ref. [14], we see that the maximum extraction current occurs at a time $t_m$ corresponding to $m \approx M/3$, regardless of the number of layers used to discretize the device, and the current maximum at this point differs by only about 0.2% from $j_m$ intervening in expression (1). Therefore, our model is consistent with previous findings [14,24] and can now be generalized to obtain cross-sectional information on the photocarrier mobility, which was not possible with non-focal illumination of the active layer, unless very arbitrary assumptions on the profile of $\mu_k$ are made.

### *3.2. Highly focal illumination of the active layer in semitransparent solar cells*

In cs-p-CELIV, also shown in Figure 3a, we are probing a device by varying the level $k$ of at which the focal plane sits inside the active layer through a piezo-scanner. Thus, cs-p-CELIV is offering a set of $f = 1\ldots M$ different values of $\Delta_{1f}$. As far as an optically absorbing semiconductor is concerned, the laser power $H_k$ (in watts) reaching the $k$-th slab, situated at a level $z_k$ below the illuminated surface, decays along $z$ according to Beer-Lambert law:

$$H_k = H_0 \exp(-\beta_0 z_k), \qquad (7)$$

where $H_0$ is the laser power at the illuminated surface and $\beta_0$ is the semiconductor's optical absorption coefficient at the wavelength of the laser, which is typically $\sim 10^4$-$10^5$ cm$^{-1}$ in the fundamental absorption photon energy range. Therefore, the amount of laser power (in watts) absorbed in the $k$-th slab between $z_k$ and $z_k + \Delta z$ [with $\Delta z = D/M$ and $z_k = (k-1)\Delta z$] is:

$$W_k = -\frac{dH_k}{dz_k}\Delta z = H_0 \beta_0 \Delta z \exp(-\beta_0 z_k). \qquad (8)$$

In a semitransparent device, for which $D \lesssim \beta_0^{-1}$, one can assume $\exp(-\beta_0 z_k) \approx 1$. This indicates that the *absolute* photon power absorbed within each semitransparent slab is constant throughout the $M$ layers of the device, where

$$W_k \approx H_0 \beta_0 \Delta z = W_0 \qquad (9)$$

irrespective of $k$, and this situation is depicted in Figure 3b.

However, the absorbed power *density* varies throughout the layers –because the beam defocuses away from the focal plane– and can be expressed as $P_k = W_k/(A_k \Delta z)$ with the area of the focal plane being the smallest, as seen in Figure 3b. By adopting the approximation of highly focal illumination, we will here assume that $A_k = A_f$ at the focal plane, with $A_k = A_0$ elsewhere, which will be relaxed later in sect. 3.3. With this approximation, one thus obtains:

$$P_k = \begin{cases} P_0 \approx W_0/(\Delta z A_0) & k \neq f \\ P_f \approx W_0/(\Delta z A_f) & k = f. \end{cases} \qquad (10)$$

As far as the photo-physics of solar cell materials is concerned, [25] the strong difference in magnitude between illuminated areas in and out of focus ($A_0/A_f \sim 10^2$ over a µm along $z$, from the PSF of our SCOM) will have profound effects on the nature of radiative recombination.

Figure 3c demonstrates that, below a critical power density, the photogenerated carrier *density* ($n_k$, in # cm$^{-3}$) is proportional to the generation rate and concentration of absorbed photons, while above this it is sublinear [25] due to a superlinear recombination rate where multimolecular recombination occurs because of neighboring electron-hole pairs. This can be summarized by

$$n_k = \begin{cases} \propto P_k/P_{\text{crit}} & k \neq f \\ \propto (P_k/P_{\text{crit}})^\varepsilon & k = f. \end{cases} \qquad (11)$$

where $\varepsilon < 1$ with, typically, $\varepsilon \approx 0.8$ in a-Si:H. [25] $P_k$ can be increased above the critical value for multimolecular recombination to appear by either i) increasing $H_0$ and $W_0$, as customarily done in the literature, [25] or ii) focussing the laser beam to $A_f \ll A_0$, as done in the present work.

The critical power density $P_{\text{crit}}$ is an intrinsic material property as it relates to the threshold distance below which electron-hole pairs recombine in a multimolecular fashion in a specific semiconductor. Nonetheless, as far as thin-film solar cells are concerned, $j_k$, the photocurrent extracted at the sandwiching electrodes, is the result of the *total* number of photocarriers extracted from each layer [i.e.: $N_k$ in # cm$^{-1}$ entering equations (2)-(4)] *not* the corresponding photocarrier density, $n_k$. This is critical in our system where $A_k$ strongly varies along the device's cross section, because the PSF decreases rapidly outside of the focal plane in a SCOM. By replacing equation (10) into (11) and rewriting the latter as a function of $N_k = n_k A_k$ in lieu of $n_k$, one obtains

$$N_k = n_k A_k = \begin{cases} N_0 \propto A_0 \times W_0/W_{\text{crit}} & k \neq f \\ N_f \propto A_0^\varepsilon A_f^{1-\varepsilon} \times (W_0/W_{\text{crit}})^\varepsilon & k = f \end{cases} \quad (12)$$

where the critical power threshold above which superlinear multimolecular recombination occurs is defined as $W_{\text{crit}} = P_{crit} \Delta z A_0$. Two different trends for $N_k$ vs. $W_k$ are predicted by equation (12) for mono- ($k \neq f$) and multi-molecular ($k = f$) recombination. These are shown in Figure 3d, on which it is worth stressing that the *absolute* photon power absorbed within each slab is constant and, therefore, $N_0$ and $N_f$ are both functions of the same abscissa, $W_0$.

Figure 3d demonstrates that $N_0 > N_f$ for any $W_0 > W_{\text{crit}}$. It shows that, as far as the total photocarrier generation is concerned, the higher the laser power the lesser the focal plane is relevant. Such a confocal selectiveness that "burns out" the plane at $k = f$ is more evident at the highest powers where, at the limit of $W_0$ (and $H_0$) tending to infinity, photocarriers chiefly originate from any slab, except the focal one. This can also be seen by taking the ratio between the two trends from equation (12), which elides the unknown proportionality coefficient:

$$r_f = \frac{N_f}{N_0} = \left(\frac{A_f}{A_0}\right)^{1-\varepsilon} \times \left(\frac{W_{\text{crit}}}{W_0}\right)^{1-\varepsilon}, \quad (13)$$

where $0 < r_f < 1$ because $A_f \ll A_0$ and $0 < 1 - \varepsilon < 1$.

The fact that, for a sufficient laser power, there is a decrease in the total number of photogenerated carriers is what allows for cross-sectional sensitivity in cs-p-CELIV experiments. Crucially, such decrease was shown to depend solely on the total power, and not the power density. We can exploit this cross-sectional sensitivity by dividing equation (4) by $N_0$, leading to

$$\mu_1 + \ldots + \mu_{f-1} + r_f \mu_f + \mu_{f+1} + \ldots + \mu_M = \frac{D}{eBN_0} \Delta_{1,f}. \tag{14}$$

where the additional subscript $f$ in $\Delta_1$ is added to signal that focal illumination at the plane $k = f$ is applied. For quantitative data analysis, $N_0$ can thus be approximated to its average as determined from a macroscopic, single point, p-CELIV measurement [13] as

$$N_0 = \frac{3\varepsilon_d}{2e\bar{\mu}} \frac{j_m}{j_0 t_m}, \tag{15}$$

where $\varepsilon_d$ is the dielectric permittivity of the active layer, and the other quantities are the same as in equation (1). A system of $M$ equations in $M$ unknowns can be derived from equation (10) as

$$(\mathbf{1} - r_f \mathbf{I}) \cdot \begin{bmatrix} \mu_1 \\ \ldots \\ \mu_M \end{bmatrix} = \frac{2D}{3B\varepsilon_d} \frac{\bar{\mu} j_0 t_m}{j_m} \begin{bmatrix} \Delta_{1,1} \\ \ldots \\ \Delta_{1,M} \end{bmatrix}, \tag{16}$$

where $\mathbf{1}$ is an $M$-by-$M$ matrix of ones and $\mathbf{I}$ is the identity matrix. This system can be solved for $\mu_1, \ldots, \mu_M$ by inverting the matrix $\mathbf{1} - r_f \mathbf{I}$, and provides a first-order framework to extract mobility profiles from photocurrent raise slopes $\Delta_{1,f}$ measured by cs-p-CELIV at varying SCOM focal planes ($f = 1, \ldots, M$), along with a macroscopic p-CELIV measurement on the same system.

### 3.3. Non-transparent solar cells and Gaussian focal illumination of the active layer

If the solar cell active layer is thick enough for the beam absorption and attenuation along $z$ to be significant, the amount of laser power deposited at any slab is no longer constant but follows equation (8). In such case, equations (14) and (16) must be modified to consider the effects of beam attenuation through the active layer. A similar calculation as the semitransparent case follows, and the total photogenerated charge carriers in each plane becomes:

$$N_k = n_k A_k = \begin{cases} N_0 \propto A_0 \times W_k/W_{\text{crit}} & k \neq f \\ N_f \propto A_0^\varepsilon A_f^{1-\varepsilon} \times (W_k/W_{\text{crit}})^\varepsilon & k = f \end{cases} \quad (17)$$

where $W_k = W_0 \exp[-\beta_0(k-1)D/M]$ accordingly to equations (8) and (9). Therefore, samples where $D \gtrsim \beta_0^{-1}$ tend to suffer from high noise when scanned at high focal plane depths and run the risk of $P_f < P_{\text{crit}}$ for high $f$, thus losing spatial sensitivity. If the condition $P_f > P_{\text{crit}}$ is still valid at the highest $z_f = (f-1)D/M$, equation (14) can be modified to take the decreasing laser power into account through the whole equation (8) in lieu of (9), which yields:

$$(\mathbf{1} - r_f \mathbf{I}) \begin{bmatrix} \mu_1 \\ \dots \\ \mu_f \\ \dots \\ \mu_M \end{bmatrix} = \frac{2D}{3B\varepsilon_d} \frac{\bar{\mu} j_0 t_m}{j_m} \begin{bmatrix} \Delta_{1,1} \\ \dots \\ \Delta_{1,f} \; e^{-\varepsilon \beta_0 (f-1)D/M} \\ \dots \\ \Delta_{1,M} \; e^{-\varepsilon \beta_0 (M-1)D/M} \end{bmatrix}. \quad (18)$$

This now accounts for beam attenuation and eliminates potential overestimations of $r_f$ and consequent underestimations of the mobilities. Equation (18) remains valid, so long as $P_k$ does not drop below the critical threshold for multimolecular recombination and spatial sensitivity.

This opacity can be taken further, by also considering that, in reality, the laser power density has a Gaussian profile, with a beam area distribution of the form:

$$A_k = A_0 - (A_0 - A_f) \exp\left[-\left(\frac{z_k - z_f}{\sigma_z}\right)^2\right]. \quad (19)$$

With this assumption, there can be a potential range of $f$'s for which $P_k > P_{\text{crit}}$ for a Gaussian-broadened focal plane, and from this a range of $N_f$ for which $N_f/N_0 < 1$. This can similarly be incorporated into equation (16) or (18), towards a more realistic model for data analysis.

## 4. Results and Discussion

Figure 4 shows the data collected in a typical cs-p-CELIV experiment at varying focal plane on sample 1. In Figure 4a, a cs-p-CELIV the current transient at a specific focal plane ($z_f =$ $z_1$, at the interface with the ITO cathode) is presented as an example from sample 1. By partially "burning out" the contributions from the focal planes at $k = f$, the bulk photo-generated charge carrier density in the entire solar cell is thus decreased, and the photo-induced resistance of the device increases. The higher the mobility of photogenerated charge carriers that are "burnt out", the smaller the increase in measurement circuit resistivity becomes. Effectively, our measurement technique becomes equivalent to measuring relative changes in the circuit's RC-time constant.

From a set of experimental current transients at $z_f = z_1, \ldots, z_M$, the maximum photocurrents ($j_{m,f}$) and times ($t_{m,f}$) are determined through an automated Matlab$^{TM}$ routine, which identifies the photo-current peaks shown in the inset of figure 4(a), from any possible global maximum elsewhere in the curve. Figures 4b and c show, respectively, examples of cross-sectional maps of $j_{m,f}$ and $t_{m,f}$ for sample 1, where the $x$-axis parallel to the solar cell surface was also scanned for 12 µm, in addition to the $z$-axis, to ensure lateral uniformity. From Figure 4c it can be noticed that $t_m$ does not have any appreciable dependence on $z_f$, which proved to be true for all samples. Conversely, panel (b) shows a significant decrease of $j_{m,f}$, which is again true for all samples. Specifically, the decrease in $j_{m,f}$ is 20% at about one third of its cross-sectional depth. This leads us to suspect high sensitivity of $j_{m,f}$ to local radiative recombination events triggered by higher photon flux deposited at $z = z_f$, for example multimolecular recombination as for equation (18).

From $j_{m,f}$ and $t_m,f$ in Figures 4(b-c), the corresponding map of $\Delta_{1,f} \approx j_m/t_m$ can be generated and is shown in Figure 4(d) for sample 1. $\Delta_{1,f}$ can be plugged into equation (18) to derive the cross-sectional mobility profile. Additional parameters required by equation (18) were derived from a macroscopic, single point, p-CELIV experiment, carried out after removing the microscope objective, which enables very high focal depth due to the coherence of the laser beam. From these measurements, an average mobility for sample 1 of $\bar{\mu} = 6$ cm$^2$ V$^{-1}$ s$^{-1}$, consistent with previous reports for a-Si:H, [27] was estimated from $t_m$ = 6.5 µs and $J_m/J_0$ = 0.0624, which were used in equation (1). As the mobility of electrons is roughly two orders of magnitude higher than holes in intrinsic a-Si:H, we assume holes are stationary for the duration of the extraction, and thus all mobilities presented are of electrons. This is a customary assumption in a-Si:H solar cell CELIV measurements.[27] Measurements on active layers with balances mobilities would require the use of block electrodes to prevent the extraction of one type of carrier. $D$ = 1.2 µm, obtained from Figure 2b, and the voltage ramp ($B$ = 2.45 $10^5$ V s$^{-1}$) applied to the function generator were also used, with a delay time of $t_d$ = 200 ns. This delay, while allowing for thermalization of photogenerated charge carriers, also allows for the more rapid multimolecular recombination to occur, leaving fewer carriers free in the focal plane for extraction by the time the ramp starts. This value has been chosen for the specific a-Si:H cells studied, however for any active layer material it can be chosen such that extraction occurs before any recombination other than multimolecular has taken place, as they will always be slower. For samples 2 and 3, average mobilities of $\bar{\mu}$=2.4 cm$^2$ V$^{-1}$ s$^{-1}$ and $\bar{\mu}$ = 3.7 cm$^2$ V$^{-1}$ s$^{-1}$ respectively, were calculated from similar macroscopic p-CELIV measurements. These average mobility values were used to determine $r_f$ appearing in equation (18) and are summarized in Table I. $r_f$ was set in a way that the average of all mobility values $\mu_1, ..., \mu_M$ was equal to the device's bulk mobility $\bar{\mu}$. Because data from all samples was shown to be very uniform along the *x*-axis, data points along *x* were averaged to improve the signal-to-noise ratio, as shown in Figure 5a-c for samples 1, 2, and 3 respectively, which reports (red dots) the values of $\Delta_{1,f} = \Delta_1(z)$ used to determine the mobility profile along the *z*-axis from equation (18).

Figures 5d-f present the cross-sectional mobility profiles along *z* for each sample, determined from $\Delta_{1,f}$ using the models derived in sections 3.2 and 3.3 under the assumptions of a semi-transparent and non-transparent solar cell, respectively. µ(*z*) for sample 1 clearly shows a nonuniform and triangular profile, with the majority carrier mobility gradually increasing from the Al anode up to three quarters of the active layer thickness, with a subsequent and sudden decrease in the proximity of ITO, possibly at the onset of the *p*-type doped a-Si:H interface with the cathode [see Figure 2a]. Self-consistency of the model used to derive the mobility profile can be tested by plugging $\mu_f$ into equation (18) and extracting a calculated value of $\Delta_{1f,Calc}$, which must favourably compare with its experimentally measured counterpart, $\Delta_{1,f,Meas}$. To this end, a hypothesis on $n_k$ to be used in equation (18) is necessary for the calculation of $P_0$. In fact, it is not true that $\Delta z \ll D$ for any of the samples studied, as the hypothesis of high cross-sectional resolution would require. On the contrary, if we assume equation (18) along with a Gaussian beam profile, a very accurate estimate of $\Delta_1(z_f)_{Calc}$ can be obtained with $P_0 = 6 \; 10^7$ W cm$^{-3}$ and $\beta_0 = 2.5 \; 10^5$ cm$^{-1}$, as shown in Figure 5(a) for sample 1. For samples 2 and 3, the laser beam intensity was lower, therefore a lower $P_0 = 7 \; 10^6$ W cm$^{-3}$ was used for simulations, leading to accurate results. Even though our calculations converge in one iteration with a standard deviation

$$STD = \sum_{f=1}^{M} \frac{|\Delta_{1,f,Calc} - \Delta_{1,f,Meas}|}{\Delta_{1,f,Calc}} \tag{20}$$

better than 5% for all samples, this can easily be extended to a multi-iteration routine by recalculating the mobility profile until a minimum threshold is reached. Although Figures 5a-c show that this iterative approach is not necessary for the reported a-Si:H devices, it may become essential for analyzing other solar cells with more varying mobility profiles. This option increases the flexibility of our mobility profile characterization tool.

A possible explanation for the triangular mobility profile across sample 1 is related to the cross-sectional variation of the hydrogen content in the a-Si:H active layer. Hydrogen is known to locally effect the mobility of a-Si:H because of its ability to passivate the coordination defects in this amorphous system.[28] While atoms in crystalline Si are all four-fold coordinated, a relatively large number (up to $10^{20}$ cm$^{-3}$ [16]) of Si sites in a-Si are three-fold coordinated, which results into dangling bond states situated at mid-gap, as shown in Figure 6a. These gap states act as traps for photogenerated electrons and holes and limit their mobility. Introduction of hydrogen during a-Si:H growth leads to the replacement of a significant proportion of dangling bonds by relatively strong SiH$_n$ groups with bonding and antibonding states sitting in the valence and conduction bands, respectively, and thus some defects become passivated.[29] Figure 6b demonstrates the role of dangling bond passivation in removing electronic states from mid-gap in a-Si:H, which may be expected to locally provide beneficial effects to the carrier mobility as mid-gap states may act as charge traps.[29] Thus, fewer charge traps and higher mobility are expected in slabs where the H content is locally higher.[30] Further, it has been shown previously that passivation of Si coordination defects via hydrogen bonds reduces the strained bonds in the amorphous silicon network [31]. This reduced strain drastically reduces the localization of electronic orbitals, primarily those in the mobility gap, and therefore increases carrier mobility.

The results of ERDA and SIMS measurements performed on the active layer of our a-Si:H solar cell are displayed in Figure 7a and confirms the predictions drawn from the variations in mobility along the solar cell profile. The SIMS and ERDA hydrogen profiles agree both in terms of atomic percentage and H distribution along the active layer profiles, with a triangular shape and decreasing H content closer to the Al anode. Figure 7b shows the hydrogen profile determined from simulating the experimental ERDA signal using the SIMNRA code.[22] It can be observed that a uniform H profile does not satisfactorily simulate the ERDA signal, while the nonuniform profile in Figure 7b offers a good simulation of the experimental data. Comparison of Figure 7b and 5b indicates that a correlation between H content and carrier mobility can be drawn.

Figure 7c shows the hydrogen content as a function of the mobility determined with a non-transparent solar cell model, as described in section 3.3. In the proximity of the Al anode, where the electron mobility is expected to be the most affected by positively charged traps. Figure 7d shows both hydrogen content and mobility ass a function of z. There is a direct correlation between H concentration and improved mobility, consistent with H passivation of electron-trapping gap states, as previously shown in Figure 6. In the proximity of the ITO cathode, where the H content is high enough, hydrogen seems to play a less significant role, and our measured mobility is independent of H content.

The picture that can be drawn from our cs-p-CELIV measurements is in excellent agreement with the commonly accepted carrier transport models of a-Si:H.[16, 30- 32] In a-Si:H solar cells, phosphorous and boron need to be incorporated at part-per-cent levels [33] to provide a suitable amount of minority and majority carriers. Virtually all of the group 13 and group 15 impurities are, in fact, trivalent and pentavalent, and, in any case, p-doped and n-doped layers are very thin (a few nm) and positioned in the proximity of the electrodes. Therefore, our cross-sectional mobility measurements are predominantly probing the majority carrier mobility within the relatively thicker intrinsic layer of the device, where an important role in controlling the mobility is played by the residual amount ($10^{17}$ cm$^{-3}$, or less [16]) of dangling bonds not passivated by hydrogen.[16] Within such layer, the dangling bond concentration is low where the H content is sufficiently high (i.e. in the bottom half of the device, closer to the ITO cathode). Therefore, in the bottom 50% of the device intrinsic region, coordination defects minimally affect the mobility that reaches the highest possible value consistent with the disordered nature of the material.

On the contrary, in the top 50% region of the intrinsic a-Si:H layer, which is closer to the Al anode, the H content in our device is significantly lower, as confirmed by both the SIMS and ERDA measurements. The concentration of dangling bonds is consequently larger in this region, where they act as traps for the majority carriers, as well as leading to more localized orbitals, both of which limit their mobility. For this reason, the mobility is expected to decrease at decreasing H content in this region, as experimentally demonstrated by Figure 7c. We can therefore conclude that, due to their excellent correlation with cross-sectional H content, our cross-sectional majority-carrier mobility measurements, are in excellent agreement with the physical understanding of the device used to validate them.

In contrast with sample 1, the mobility profile obtained for sample 2 appears much more symmetric; the mobility peaks in the middle of the active layer and decreases equally towards each electrode. This mobility is less likely to be caused by variations in hydrogen profile, and more likely to be caused by doped layers. Due to a non-zero diffusion coefficient of boron and phosphorous in a-Si:H, there will be a gradient of these dopants near the doped layers in the pin architecture [34]. Dopant concentration is known the increase electron scattering, and therefore decrease mobility by an amount proportional to $(1-e^{-\rho/d})$ where $\rho$ is the free electron density, and d is the dopant density [35]. Therefore, as the impurity density increases near electrodes due to diffusion, there is an increase in electron scattering which decreases the mobility.

The photocarrier mobility profile measured from sample 3 also has a very symmetric shape. Due to the thinness of this sample, we see minimal effect due to dopant diffusion as in sample 2, and no indications of variations in hydrogen content as in sample 1. These perhaps offer insight into why this cell has the highest mobility, as seen in Table I. Despite having a lower maximum mobility than sample 1, which has a comparable thickness, charges are extracted from each layer identically, showing that consistent mobility in each plane is critical for device performance.

While our cross-sectional mobility profile measured by cs-p-CELIV correlates extremely well with the hydrogen profile detected by ERDA, it is worth noting that H profiling in a-Si:H requires sophisticated ion measurements. Conversely, our cs-p-CELIV benchtop system is suitable to work in air and is simple to implement at low budget costs. Furthermore, cs-p-CELIV is a non-destructive technique and, differently from vacuum-related characterization methods, it is suitable to rapid quality control in large commercial panels made of thin-film solar cells. This is also witnessed by the possibility to perform line scans along the *x*-axis, as shown in Figure 4.

There are a few limitations of our method to consider, one of the more obvious being the sample thickness, as it relates to the axial resolution of the confocal microscope's illumination function. This resolution is less strict than for typical confocal imaging, as collecting in focus photons is not the goal, but rather the thickness of the cross-sectional plane in which multimolecular recombination occurs must be sufficiently thinner than the total device thickness to achieve true spatial resolution. The maximum thickness is also limited when considering non-transparent samples, as the laser pulse intensity decays exponentially throughout the device due to absorption. If the intensity decays within the active layer such that $N_k \ll \mu_k eB/D$, the resulting variation of $\Delta_{1f}$ will be imperceptible for all planes f > i. If $\delta j$ is the minimum current change that can be detected, then $\delta j$ will not be noticeable in samples with mobility less than or close to $D/(MeBt_m)$ and will lead to drastic underestimation of the mobility variation. There is, however, no expected limitation on a maximum measured mobility, as our model and experimental design make no assumptions on low mobility. One final restriction implicit in our model is the ratio of electron and hole mobilities. We assume that one more mobile species is fully extracted before the other species begins to react to the applied electric field, and thus in all layers between the extraction layer and the extracting electrode ($l(t)<z<D$) the densities of electrons and holes is equal, and no additional electric field is created to influence the unextracted charge carrier's behaviour.

## 4. Conclusion

In conclusion, the integration of p-CELIV measurements with a SCOM has allowed for spatially dependent measurements of the majority charge carrier mobility along the $z$-axis of micrometer-thick solar cells, by exploiting enhanced multimolecular recombination at the SCOM focal plane. One of the mobility profiles that was obtained correlates well with the hydrogen profile of the a-Si:H active layer, and adds insight into the role of hydrogen in the optoelectronic properties of a-Si:H. The more symmetric mobility profiles were explained by considering the role of the depletion layer formed at a metal-semiconductor interface. While our technique was demonstrated on an a-Si:H device, it is widely applicable to a wide range of thin-film photovoltaic devices.[36] The dominance of multimolecular recombination at high illumination is a general property of a vastity of solar cell materials, and therefore we are not restricted by sample. As we have demonstrated, the technique we developed allows for unprecedented cross-sectional investigations into the devices' structure. In summary, cs-p-CELIV is a powerful non-destructive technique and, differently from vacuum-related characterization methods such as SIMS and ERDA, it is suitable to rapid quality controls in large commercial panels made of thin-film solar cells, in which reproducible and carefully engineered mobility profiles across the active layer are vital.

**Acknowledgements**

We thank Dr. T. Simpson and Mr. T. Goldhawk of Western's Nanofabrication Facility for technical support with FIB, SEM and EDX and Dr. L. Goncharova and Mr. J. Hendriks of Interface Science Western as well as Dr. M. Biesinger and Mr. G. Good of Surface Science Western for their assistance with ERDA and SIMS, respectively. This study was supported by the Canada Research Chair secretariat, the Canada Foundation for Innovation, and the Natural Sciences and Engineering Research Council of Canada (NSERC) under the Discovery Grant program.

**Data Availability Statement**

The data that support the findings of this study are available from the corresponding author upon reasonable request.

# References

1. A.V. Shah, R. Platz, H. Keppner, Thin-film silicon solar cells: A review and selected trends, Solar Energy Materials and Solar Cells 38, 501-520(1995)

2. P. Roy, N. K. Sinha, S. Tiwari, A. Khare, A review on perovskite solar cells: Evolution of architecture, fabrication techniques, commercialization issues and status, Solar Energy 198 (2020) 665-688

3. M. Green, Thin-film solar cells: review of materials, technologies and commercial status, Journal of Materials Science: Materials in Electronics 18, 15-19(2007)

4. W. Tress, K. Leo, M. Riede, Optimum mobility, contact properties, and open-circuit voltage of organic solar cells: A drift-diffusion simulation study, Phys. Rev. B 85, 155201(2012)

5. A. Privkas, N. Sariciftci, G. Juška, R. Österbacka, A Review of Charge Transport and Recombination in Polymer/Fullerene Organic Solar Cells, Prog. Photovolt: Res. Appl. 15, 677-696(2007)

6. E.A. Schiff, Low-mobility solar cells: a devie physics primer with application to amorphous silicon, Solar Energy Materials and Solar Cells 78, 567-595(2003)

7. G. Chen, Y. Chen, C. Lee, H. Lee, Performance improvement of perovskite solar cells using electron and hole transport layers, Solar Energy 174, 897-900(2018)

8. J. Shieh, C. Liu, J. Meng, S. Tseng, Y. Chao, S. Horng, The effect of carrier mobility in organic solar cells, J. Appl. Phys. 107, 084503(2010)

9. P. LeComber, W. Spear, D. Allan, Transport studies in doped amorphous Silicon, J. Non-Cryst. 32, 1-15(1979)

10. W. Spear, The Hole Mobility in Selenium, Proc. Phys. Soc. 76, 826(1960)

11. D. Goldie, The determination of carrier lifetimes and associated mobility magnitudes using photoconductivity recovery dynamics in thin-film amorphous semiconductors, Thin Solid Films 675, 11-15(2019)

12. G. Juška, M. Viliūnas, K. Arlauskas, N. Nekrašas, N Wyrsch, L. Feitknecht, Hole drift mobility in μc-Si:H, J. Appl. Phys. 89, 4971(2001)

13. G. Juška, K. Arlauskas, N. Nekrašas, J. Stuchlik, X. Niquille, N. Wyrsch, Features of charge carrier transport determined from carrier extraction current in μc-Si:H, J. Non-Cryst. Solids, 299-302, 375-379(2002)

14. G. Juška, K. Genevičius, M. Viliunas, K. Arlauskas, H. Stuchlíková, A. Fejfar, J. Kočka, New method of drift mobility evaluation in μc-Si:H, basic idea and comparison with time-of-flight, J. Non-Cryst. Solids, 266-269, 331-335(2000)

15. A. Ashraf, D. M. N. M. Dissanayake, M. D. Eisaman, Measuring charge carrier mobility in photovoltaic devices with micron-scale resolution, Appl. Phys. Lett. 106, 113504(2015)

16. R. A. Street, 1991, Hydrogenated Amorphous Silicon, Cambridge University Press, Cambridge, UK.

17. G. Juška, N. Nekrašas, V. Valentinavičius, P. Meredith, A. Pivrikas, Extraction of photogenerated charge carriers by linearly increasing voltage in the case of Langevin recombination, Phys. Rev. B 84, 155202(2011)

18. B. Tremolet de Villers, C. Tassone, S. Tolbert, B. Schwartz, Improving the Reproducibility of P3HT:PCBM Solar Cells by Controlling the PCBM/Cathode Interface, J. Phys. Chem. Lett. C 113, 18978-18982(2009)

19. S. Venkatesan, E. Ngo, D. Khatiwada, C. Zhang, Q. Quiao, Enhanced Lifetime of Polymer Solar Cells by Surface Passivation of Metal Oxide Buffer Layers, ACS Appl. Mater. Interfaces 7(29), 16093-16100(2015)

20. A. Mozer, N. S. Saricifti, Charge transport and recombination in bulk heterojunction solar cells studied by the photoinduced charge extraction in linearly increasing voltage technique, Appl. Phys. Lett. 86, 112104(2005)

21. S. Fonash, 2010, Solar Cell Device Physics, second ed., Academic Press, Burlington, MA

22. S.M. Mayer, Improved physics in SIMNRA 7, Nuclear Instruments and Methods in Physics Research B 331, 176-180(2014)

23. K. Emery, Photovoltaic Callibrations at the National Renewable Energy Laboratory and Uncertainty Analysis Following the ISO 17025 Guidelines, National Renewable Energy Laboratory (NREL), Golden, CO, NREL/TP-5J00-66873, 2016.

24. G. Juška, K. Arlauskas, M. Viliūnas, J. Kočka, Extraction current transients: New Method of Study of Charge Transport in Microcrystalline Solids, Phys. Rev. Lett. 84, 4946-4949 (2000)

25. W. Fuhs, Geminate and non-geminate recombination in a-Si:H: old questions and new experiments, J. Optoelectron. Adv. M. 7, 1889-1897(2005).

26. G. D. Cody, T. Tiedje, B. Abeles, B. Brooks, and Y. Goldstein, Disorder and the Optical-

Absorption Edge of Hydrogenated Amorphous Silicon, Phys. Rev. Lett. 47, 1480-1483 (1981)

27. M. T. Neukom, N. A. Reinke, B. Ruhstaller, Charge extraction with linearly increasing voltage: A numerical model for parameter extraction, Solar Energy 85, 1250-1256 (2011)

28. G. Hahn, P. Geiger, D. Sontag, P. Fath, E. Bucher, Influence of hydrogen passivation on majority and minority charge carrier mobilities in ribbon silicon, Solar Energy Materials and Solar Cells 74, 57-63 (2002)

29. K. Gmucová, V. Nádaždy, R. Durny, The nature of mobile Hydrogen in a-Si:H-Electrochemical studies, Solar Energy 80, 694-700(2006)

30. E. Cartier, J. H. Stathis, D. A. Buchanan, Passivation and depassivation of Silicon dangling bonds at the Si/$SiO_2$ interface by atomic Hydrogen, Appl. Phys. Lett. 63, 1510(1993)

31. R. V. Meidanshahi, S. Bowden, S. M. Goodnick, Electronic structure and localized states in amorphous Si and hydrogenated amorphous Si, Phys. Chem. Chem. Phys., 21, 13248(2019)

32. R.A. Street, Recombination in a-Si:H: Defect luminescence, Phys. Rev. B 21(12), 5775-5784 (1980)

33. W.E. Spear, P.G. LeComber, Substitutional doping of amorphous silicon, Solid State Communications 17(9), 1193-1196 (1975)

34. R. A. Street, C. C. Tsai, J. Kakalios, W.B. Jackson, Hydrogen diffusion in amorphous silicon, Phil. Mag. B 56(3), 305-320(1987)

35. S. S. Li, W. R. Thurber, The dopant density and temperature dependence of electron mobility and resistivity in n-type silicon, Solid-State electronics 20, 609-616 (1977)

36. M. Green, Third generation photovoltaics: advanced solar energy conversion, Springer, New York, 2003.

**Figure Captions**

**Figure 1**- (a) Diagram of p-CELIV combined with SCOM to form a cs-p-CELIV setup, where a 527-nm laser beam is pulsed through the pinhole, and then focused onto the focal plane $z_f$, which is varied by using the piezo-scanner on which the solar cell is positioned. (b) Schematic of p-CELIV experiment. Maximum extraction current occurs at $t_m$.

**Figure 2**- Thin-film a-Si:H based *p-i-n* solar cell used as test sample in this work, as well as the need for their cross-sectional investigation. (a) diagram of device architecture compared to (b) a cross-sectional SEM image of the device. (c) Cross-sectional EDX imaging in the same area as in panel b for Al and (d) Si composition to identify the electrodes and a-Si:H active layer for enhanced interpretation of subsequent cs-p-CELIV investigations.

**Figure 3**- (a) p-CELIV model for uniform active layer, based on a multi-slab discretization of the device, where physical quantities appearing in the text are indicated. (b) Schematic of illumination in a SCOM where mono- and multimolecular recombination of excess carriers at the focal plane is presented. (c) Variations of $n_k$ with laser power density, where high power density leads to a sublinear dependence on $P_k$ in the focal plane. (d) $N_k$ (green) and $N_k/N_0$ (purple) as they depend on incident power $W_0$, which decreases beyond $W_{crit}$.

**Figure 4**- (a) Typical p-CELIV current transients with SCOM confocal plane set at both a high mobility plane and a low mobility plane of the a-Si:H solar cell active layer, as well as the bulk p-CELIV curve. Inset shows zoom of photocurrent peak. (b) Photocurrent maximum ($j_{m,f}$) as a function of $x$ and $z_f$. The minimum of $j_{m,f}$ at $z_f \approx 2$ μm is situated at about three-quarters of the a-Si:H layer thickness. (c) Maximum transient time, $t_{m,f}$ as a function of $x$ and $z_f$. Differently from $j_{m,f}$, the value of $t_{m,f}$ has little dependence on $z_f$. (d) Photocurrent transient ramp $\Delta_1(x, z_f)$ that also presents a minimum at about three-quarters of the a-Si:H layer thickness, due to the minimum of $j_{m,f}(z_f)$. $\Delta_1(x, z_f)$ is then averaged along $x$, and used to determine the mobility at each $z$-axis slab.

**Figure 5-** (a-c) Measured $\Delta_1(x, z_f)$ averaged over each $x$-line for samples 1, 2, and 3 respectively. per focal plane (dots) and calculated $\Delta_1(z_f)$ (solid lines, with semitransparent and nontransparent approximations considered). It can be observed that the calculated values match the measured one indicating that the mobility values are accurate, with a less than 2% standard deviation. (d-f) Mobility profiles for samples 1, 2, and 3 respectively from eq. (18) showing a triangular profile for sample 1, both in a semitransparent (sect 3.2) and nontransparent (sect 3.3) solar cell models, and more symmetric profiles for samples 2 and 3. Dashed lines are visual aids.

**Figure 6-** (a) Random a-Si:H network, with dangling bonds represented by dots. (b) Dangling bond states lead to charge traps at mid-gap, which significantly affect the carrier mobility. A schematic of the a-Si:H density-of-states of is shown by solid lines, while dashed lines in the gap are represent traps states that are eliminated via H passivation of dangling bonds.

**Figure 7-** (a) ERDA hydrogen profile and comparison with SIMNRA simulation used to determine the H atomic percentage. It can be noticed that a uniform H profile is not able to account for the measured ERDA spectrum at the lowest channel energies. (b) H profile along the a-Si:H active layer, showing significant variations as obtained from ERDA (bars) and SIMS. (c) Correlation between H content and carrier mobility, showing an increasing majority ($n$-type) carrier mobility at increasing H content due to dangling bond passivation for the bottom 50% of the $z$-axis profile. This is consistent with data obtained from both ERDA and SIMS. Majority ($n$-type) carrier mobility is not dependent on the H content in the top 50% of the $z$-axis, where a closer correlation with the minority ($p$-type) mobility would be expected. (d) H content vs. mobility, illustrating their correlation along z, which illustrates their correlation throughout the device active layer.

**Figure 1**

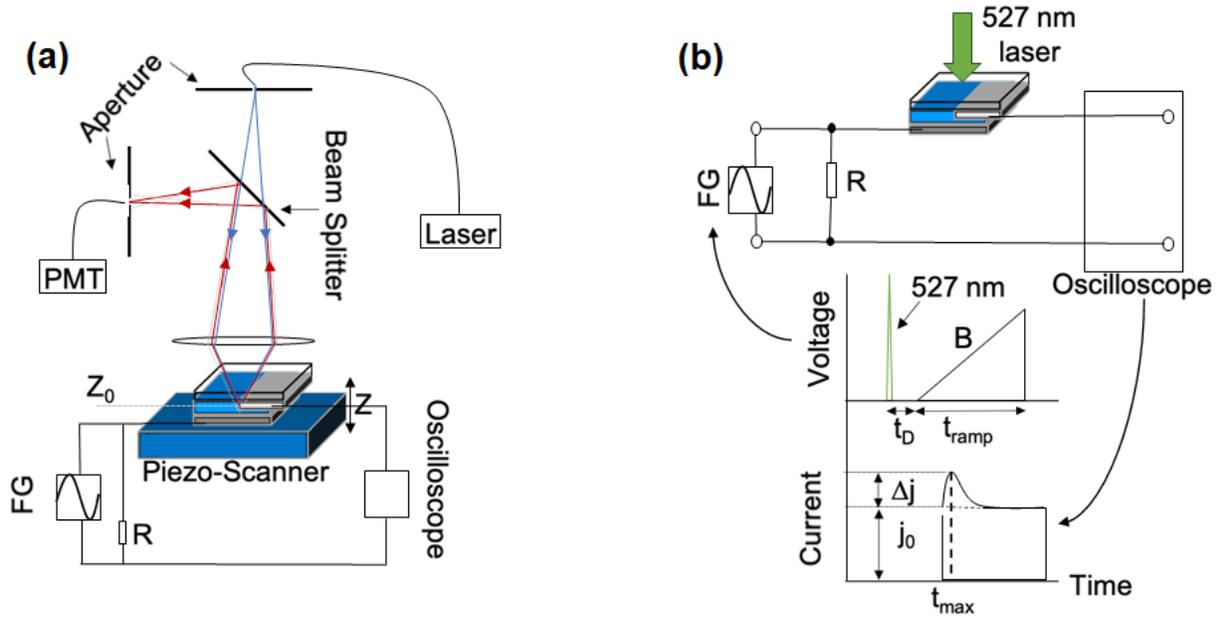

**Figure 2**

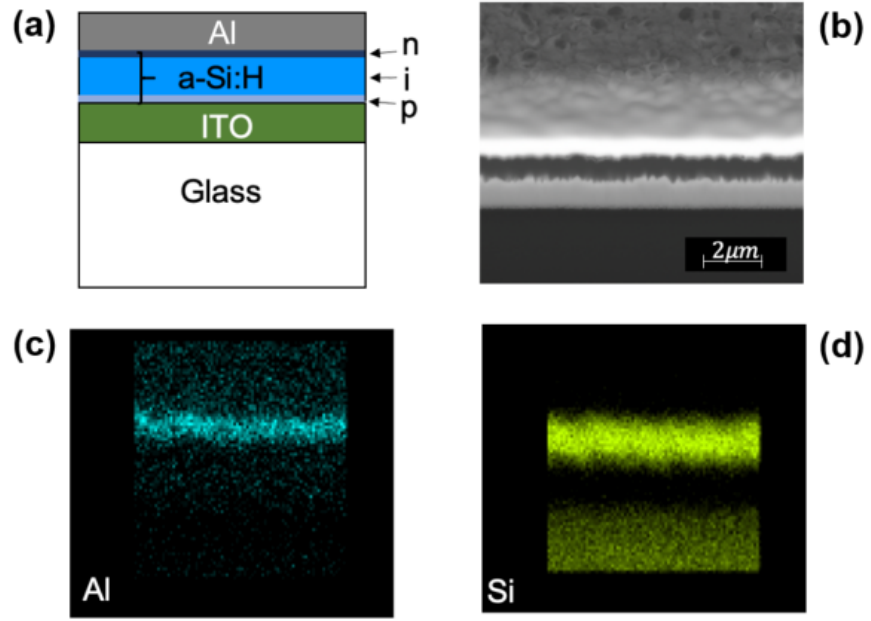

**Figure 3**

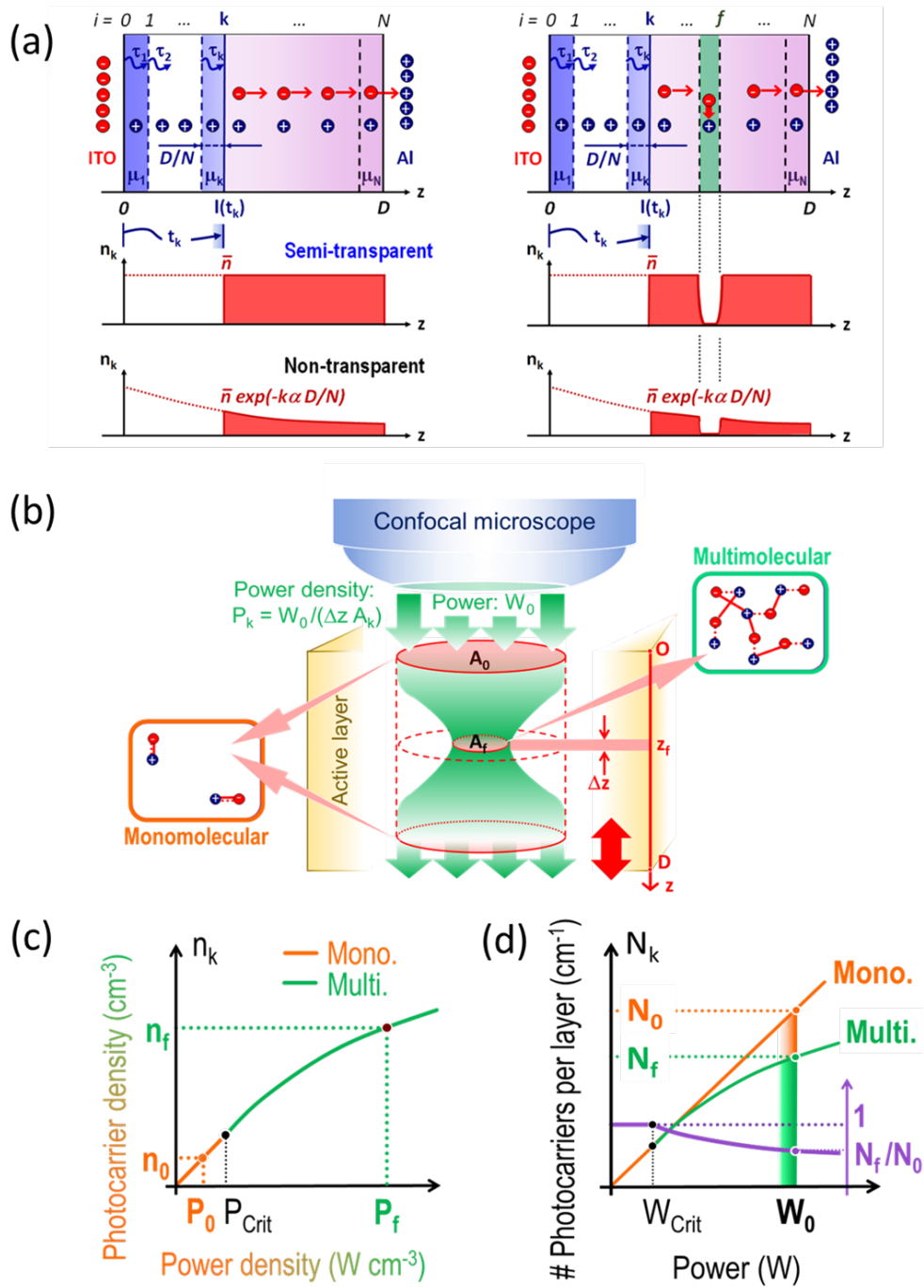

**Figure 4**

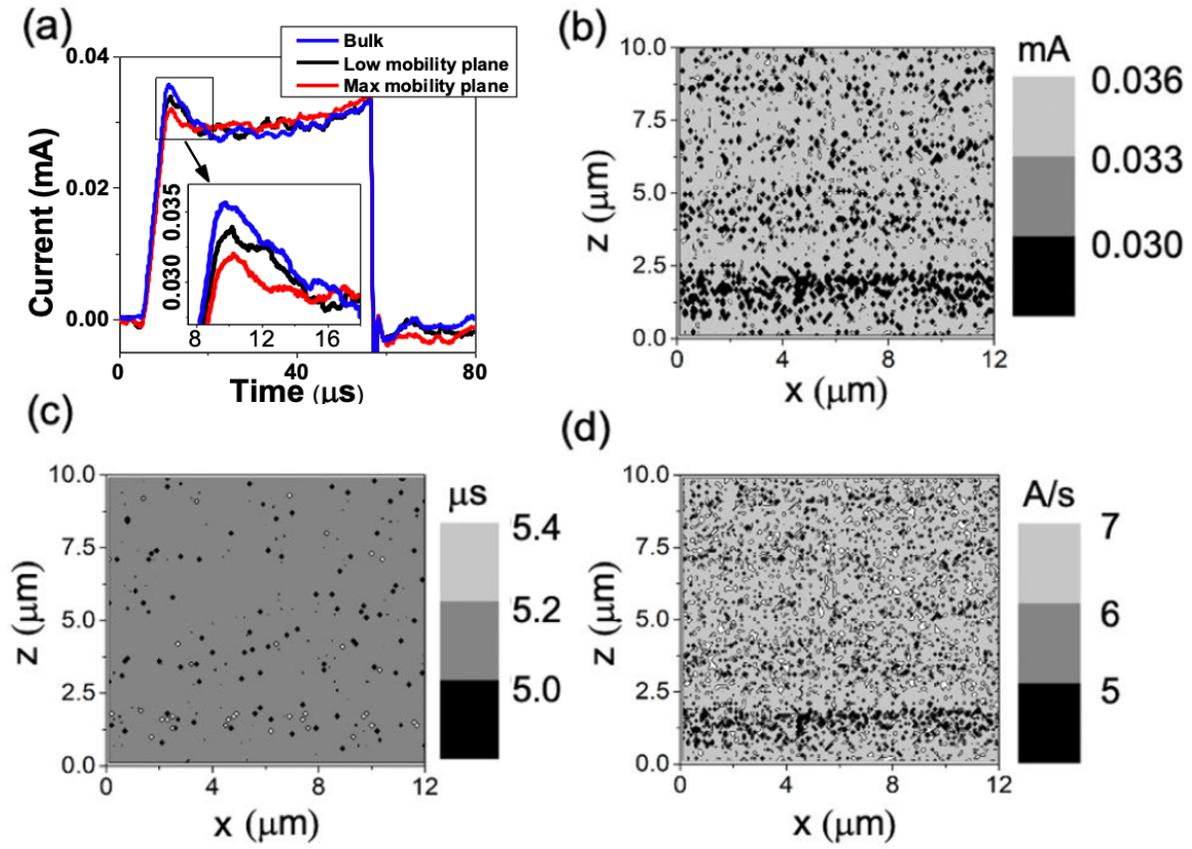

**Figure 5**

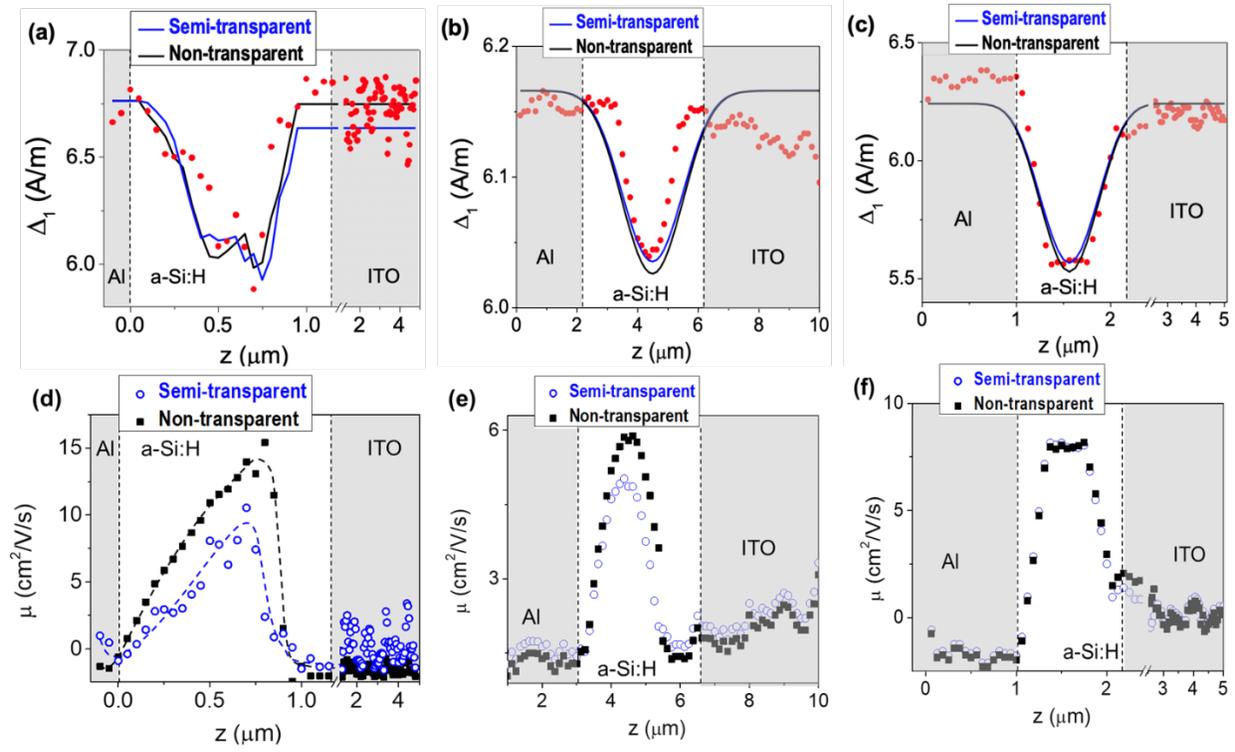

**Figure 6**

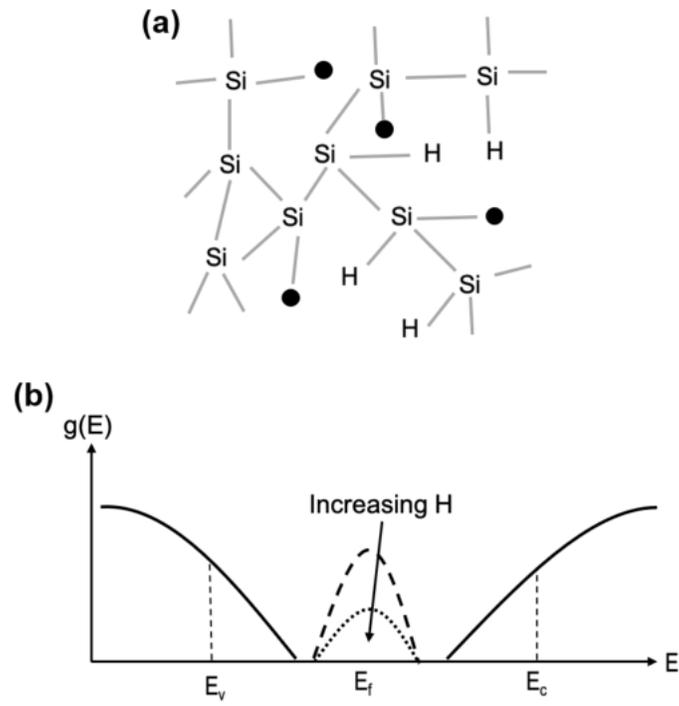

**Figure 7**

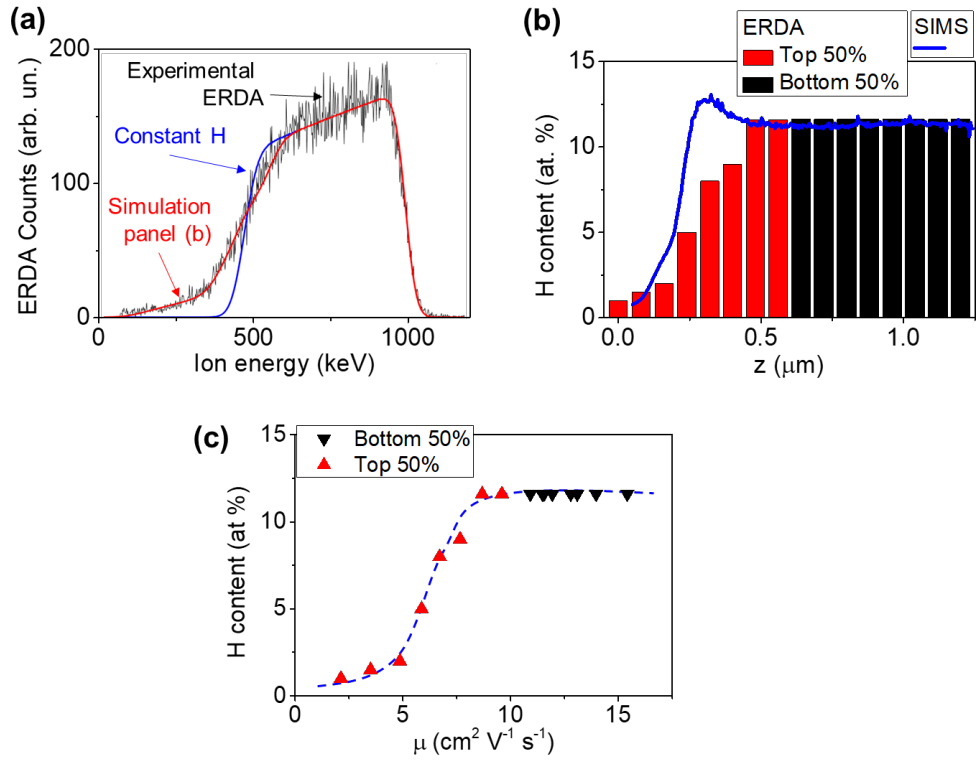

**Table I**

The surface area, short-circuit current density, open-circuit voltage, maximum power, efficiency, and $r_f$ values for all samples, all calibrated to NREL reference sample. Samples of 1 and 2 have a lower efficiency which is primarily explained by their thicker active layers, while the thinner sample 3 is of relatively higher efficiency.

|  | Area (cm$^2$) | $J_{sc}$ (mA/cm$^2$) | $V_{OC}$ (V) | $P_{max}$ (mW) | η (%) | $r_f$ |
|---|---|---|---|---|---|---|
| Sample 1 | 3.5 | 4.64 | 0.86 | 20.93 | 5.98 | 0.766 |
| Sample 2 | 6.0 | 0.80 | 1.96 | 26.46 | 4.41 | 0.795 |
| Sample 3 | 3.0 | 5.42 | 0.86 | 29.61 | 9.87 | 0.669 |